\documentclass[reprint,amsmath,amssymb,amsbsy,prb,footinbib]{revtex4-2}
\usepackage{graphicx,color}
\usepackage{braket}
\usepackage{bm}
\usepackage{color}
\usepackage{ulem}
\usepackage{verbatim}
\usepackage{times}
\usepackage[colorlinks=true, citecolor=blue, linkcolor=blue]{hyperref}

{\par\noindent{\em #1:\ }}%
{~\rule{2mm}{2mm}\par\bigskip}

\begin{document}
\title{Evolution of flat bands in two-dimensional fused pentagon network} 

\author{Tomonari Mizoguchi}
\affiliation{Department of Physics, University of Tsukuba, Tsukuba, Ibaraki 305-8571, Japan} 
\email{mizoguchi@rhodia.ph.tsukuba.ac.jp}
\author{Mina Maruyama}
\affiliation{Department of Physics, University of Tsukuba, Tsukuba, Ibaraki 305-8571, Japan} 
\author{Yasuhiro Hatsugai}
\affiliation{Department of Physics, University of Tsukuba, Tsukuba, Ibaraki 305-8571, Japan} 
\author{Susumu Okada}
\affiliation{Department of Physics, University of Tsukuba, Tsukuba, Ibaraki 305-8571, Japan} 

\date{\today}
\begin{abstract}
Theoretical quest of flat-band tight-binding models usually relies on lattice structures 
on which electrons reside.
Typical examples of candidate lattice structures include the Lieb-type lattices and the line graphs. 
Meanwhile, there can be accidental flat-band systems that belong to neither of 
such typical classes
and deriving flat-band energies and wave functions for such systems is not straightforward. 
In this work, we investigate the characteristic band structure for the tight-binding model on a network composed of pentagonal rings,
which is inspired by the theoretically-predicted carbon-based material. 
Although the lattice does not belong to conventional classes of flat band models, 
the exact flat bands appear only for fine-tuned parameters.
We analytically derive the exact eigenenergies and eigenstates of the flat bands.
By using the analytic form of the Bloch wave function, we construct the corresponding Wannier function and reveal its characteristic real-space profile. 
We also find that, even away from the exact flat-band limits, the nearly flat band exists near the Fermi level for the half-filled systems, 
which indicates that the present system will be a suitable platform for questing flat-band-induced correlated electron physics if it is 
realized in the real material. 
\end{abstract}

\maketitle
\section{Introduction}
Flat bands are one of the characteristic dispersion in crystalline systems, 
where the energy eigenvalue is constant in the entire Brillouin zone.
In flat-band systems, the diverging density of states makes the electron-electron correlation prominent. 
Indeed, exotic correlation-induced phenomena, such as ferromagnetism~\cite{Mielke1991,Tasaki1992}, superconductivity~\cite{Imada2000,Peotta2015,Aoki2020,Huhtinen2022}, 
Bose-Einstein condensation~\cite{Huber2010,Katsura2021}, and 
the zero-field fractional quantum Hall effect (i.e., the fractional Chern insulators)~\cite{Tang2011,Sun2011,Neupert2011},
have been pursued in flat-band systems.
Recently, from the experimental sides, 
variety of the flat-band systems have been found or artificially fablicated.
Examples include the twisted bilayer graphene~\cite{Cao2018,Cao2018_2,Yankowitz2019,Park2021}, 
the kagome metals~\cite{Ortiz2019,Yin2019,Kang2020,Sun2022,Ye2024},  
the molecular-based materials~\cite{Kumar2018,Shuku2023,Nemoto2024},
the cold atoms~\cite{Taie2015,Ozawa2017},
and the photonic systems~\cite{Xia2018,Leykam2018,Tang2020,Yang2024}.

From the viewpoint of the fundamental theory, 
two typical lattice structures for constructing tight-binding Hamiltonians hosting flat bands are known.
One is the sublattice-number-imbalanced bipartite (i.e., the Lieb-type) lattices~\cite{Sutherland1986,Lieb1989},
and the other is the line graphs~\cite{Mielke1991} and their variants~\cite{Miyahara2005,Lee2019,Ogata2021,Liu2022,Mizoguchi2023}.
In both of these two types of lattices, 
the formation of localized states due to the 
destructive interference mechanism plays an essential role;
such localized states turn to the flat bands in the momentum space representation, and 
they are dubbed as the compact localized states (CLSs)~\cite{Zhitomirsky2004,Bergman2008}.
Additionally, for these classes of models and their extensions, 
analytical solutions of the flat-band states can be obtained in a systematic manner~\cite{Mizoguchi2019_star,Lee2019,Mizoguchi2021,Heo2023,Marques2023}
even though there are multiple flat bands with different energies.

\begin{figure}[b]
\begin{center}
\includegraphics[clip,width = 0.9\linewidth]{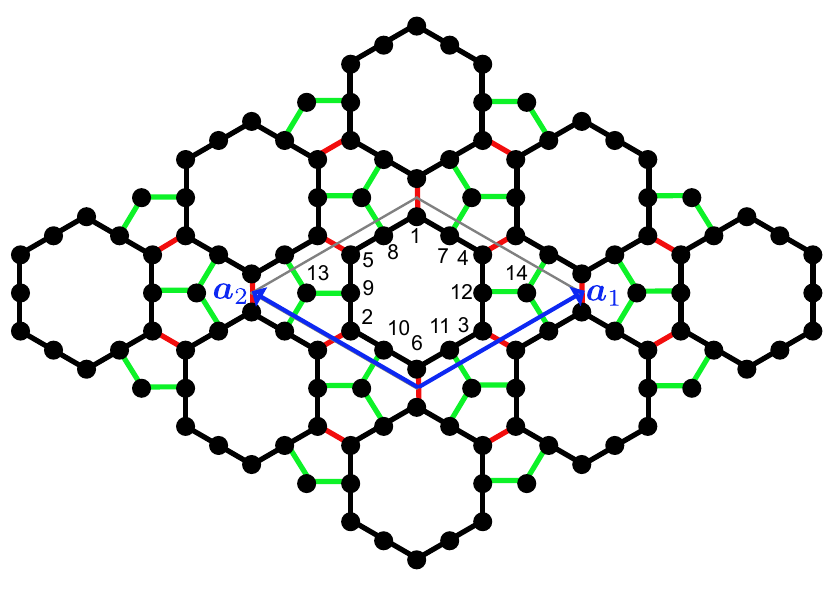}
\vspace{-10pt}
\caption{
Lattice structure considered in this paper. 
Black, red, and green bonds correspond to the transfer integrals $t_1$, $t_2$, and $t_3$, respectively. 
The lattice vectors are $\bm{a}_1 = \left( \frac{\sqrt{3}}{2},\frac{1}{2}\right)$ and $\bm{a}_2 = \left( -\frac{\sqrt{3}}{2},\frac{1}{2}\right)$.}
\label{fig:lattice}
 \end{center}
 \vspace{-10pt}
\end{figure}

Meanwhile, not all the flat-band models belong to the above two classes, 
namely, flat bands can appear accidentally. 
In such cases, in general, there is no systematic method 
to obtain analytic solutions of the flat bands.
Here we consider one of such examples.
The lattice structure we study in this paper is depicted in Fig.~\ref{fig:lattice}, 
which is a tight-binding model for the carbon allotrope 
proposed by two of the authors~\cite{Maruyama2013,Maruyama2014}
~\footnote{Note that the calculated lattice structure (i.e., the optimal position of the carbon atom) is slightly different from Fig.~\ref{fig:lattice},
but the connectivity of the bonds are equivalent.}. 
The proposed material is of interest in terms of various applications
such as H$_2$ separation~\cite{Zhu2015}, the anode for Li ion batteries~\cite{Zeng2020}, and desalination~\cite{Niu2023}.
The lattice geometry of Fig.~\ref{fig:lattice} is characteristic in that it consists of a network of fused pentagons.
By investigating the evolution of the band structures upon changing the parameters,
we find that there are two limiting cases that have exact flat bands.
Interestingly, the origin of the flat band in these two cases are quite different, namely, 
one is due to the destructive interference effect and the other is attributed to the accidental reason.
For both of these two limits, 
we derive the exact flat-band solutions, and elucidate that the flat-band wave functions are 
completely different to each other. 
We also find that the nearly flat bands survive even away from the exact flat-band limits,
which indicates that the strongly-correlated physics will be feasible in the real materials hosting this network of fused pentagons
even though the transfer integrals of the real materials is away from the ideal flat-band limit.

The rest of this paper is structured as follows.
In Sec.~\ref{sec:model}, we introduce our tight-binding model,
which is inspired by the carbon-network material~\cite{Maruyama2013}.
Our main results are presented in Sec.~\ref{sec:band},
where we show the band-structure evolution upon changing the hopping parameters.
The real-space pictures for two exact-flat-band limits are also presented. 
The summary of this paper is presented in Sec.~\ref{sec:summary}.
In Appnedix~\ref{app:mo}, we provide a detailed explanation about the flat band for $t_3 = 0$.
In Appendix~\ref{app:FB_exact}, we explain the derivation of the analytic solution of the flat bands for $t_1 = t_2 = t_3 $. 
Finally, in Appendix~\ref{app:topo}, we show that we can construct the topological flat band model by 
introducing the complex next-nearest-neighbor hoppings.

\begin{figure*}[tb]
\begin{center}
\includegraphics[clip,width = 0.95\linewidth]{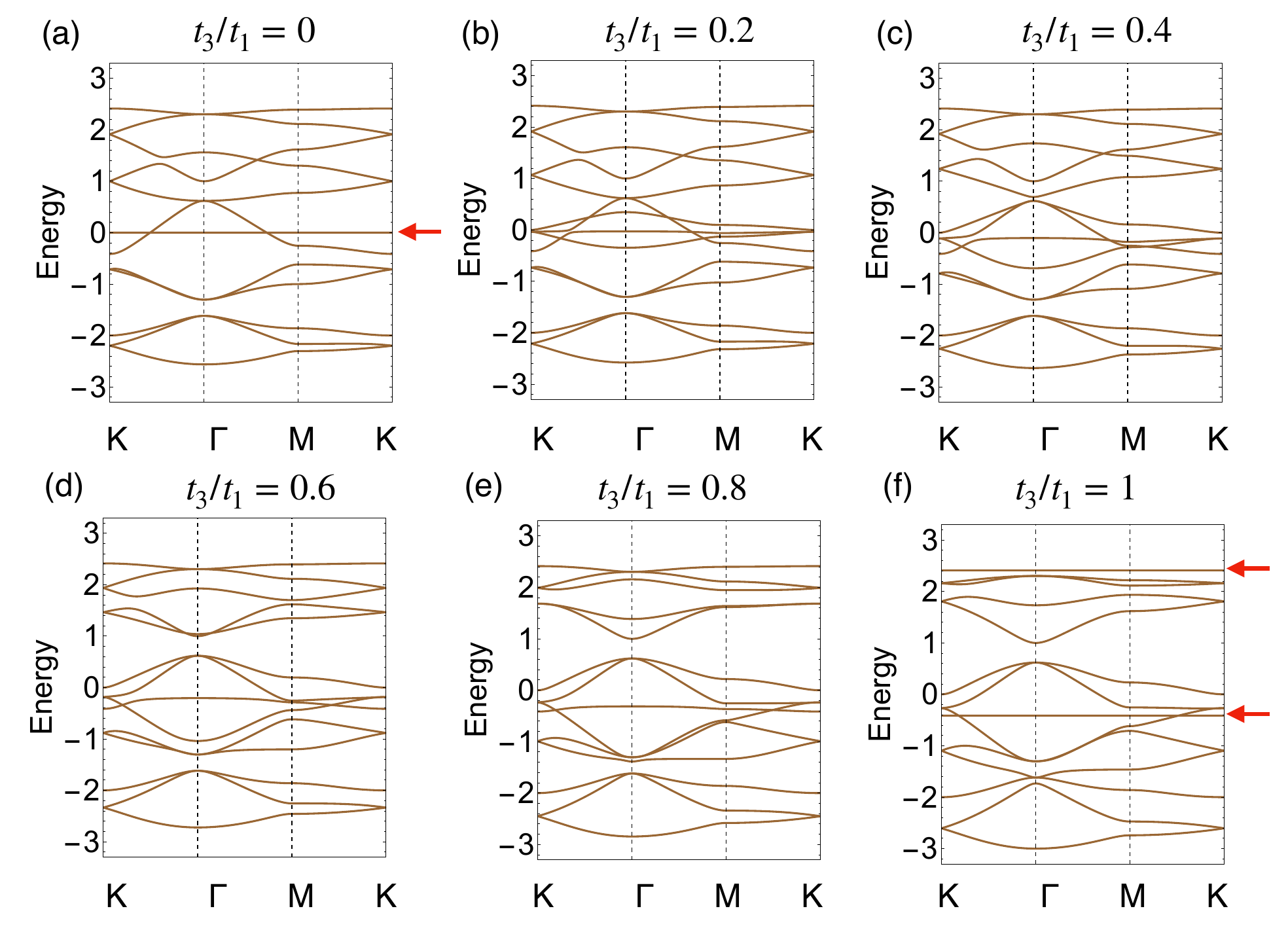}
\vspace{-10pt}
\caption{
Band structures of the model of Eq.~(\ref{eq:Ham}).
We fix $t_1 = t_2 = -1$, and vary $t_3/t_1$ as indicated in each panel. 
The high-symmetry points in the Brillouin zone are 
$\Gamma = (0,0)$, K$ = \left(\frac{2\pi}{\sqrt{3}}, \frac{2\pi}{3} \right)$, and M$ = \left(\frac{\pi}{\sqrt{3}}, \pi \right)$.
Red arrows in the panels (a) and (f) point the exact flat bands.
}
  \label{fig:band}
 \end{center}
 \vspace{-10pt}
\end{figure*}
\section{Model \label{sec:model}}
We consider the tight-binding model defined 
on a lattice of Fig.~\ref{fig:lattice} for spinless, single-orbital fermions.
There are 14 sites per unit cell. 
In the real space, the Hamiltonian reads 
$H = \sum_{\langle i,j \rangle} t_{i,j} c^\dagger_{i} c_j + \left( \mathrm{h.c.}\right)$,
where $c_j$ represents the annihilation operator of the fermion at the site $j$ and $t_{i,j}$ represents 
the transfer integral between the sites $i$ and $j$.
There are three types of bonds, which we denote by black, red, and green bonds in Fig.~\ref{fig:lattice}. 
Performing the Fourier transform, we obtain the momentum-space representation of the Hamiltonian 
$H = \sum_{\bm{k}} \bm{c}^\dagger_{\bm{k}} \mathcal{H}_{\bm{k}}   \bm{c}_{\bm{k}}$
where $c_{\bm{k}} = \left( c_{\bm{k},1}, \cdots, c_{\bm{k},14} \right)^{\rm T}$ 
and $\mathcal{H}_{\bm{k}}$
is the Bloch Hamiltonian in the form of $14 \times 14$ matrix:
\begin{align}
\mathcal{H}_{\bm{k}} = 
\begin{pmatrix}
\mathcal{O}_{3,3} & S_{\bm{k}} & T_1 & \mathcal{O}_{3,2}\\
S_{\bm{k}}^\dagger &\mathcal{O}_{3,3} & T_2 & \mathcal{O}_{3,2}  \\
T_1^\dagger & T_2^\dagger & \mathcal{O}_{6,6} & T_{\bm{k},3} \\
 \mathcal{O}_{2,3}&  \mathcal{O}_{2,3} &  T^\dagger_{\bm{k},3}  & \mathcal{O}_{2,2} \\
\end{pmatrix}, \label{eq:Ham}
\end{align}
where 
\begin{align}
S_{\bm{k}}  = 
\begin{pmatrix}
0 & 0 & t_2 e^{i (\theta_{\bm{k},1} + \theta_{\bm{k},2} ) } \\
t_2 e^{-i \theta_{\bm{k},1}} & 0 & 0 \\
0 & t_2e^{-i \theta_{\bm{k},2} } & 0 \\
\end{pmatrix}, \label{eq:smat}
\end{align}
\begin{align}
T_1 = \begin{pmatrix}
t_1 & t_1 & 0 & 0 & 0& 0 \\
0 & 0 & t_1 & t_1 & 0 & 0\\
 0 & 0 & 0& 0 & t_1 & t_1\\
\end{pmatrix}, \label{eq:T1}
\end{align}
\begin{align}
T_2 = \begin{pmatrix}
t_1 &0 & 0 & 0 & 0&  t_1 \\
 0 & t_1 & t_1 & 0 & 0 & 0 \\
 0 & 0 & 0&  t_1 & t_1 &0 \\
\end{pmatrix},\label{eq:T2}
\end{align}
and 
\begin{align}
T_{\bm{k},3} = 
\begin{pmatrix}
 t_3e^{i\theta_{\bm{k},1}} & 0 \\
 0 & t_3e^{i\theta_{\bm{k},2}} \\
 t_3 & 0 \\
 0 & t_3e^{-i \theta_{\bm{k},1} } \\
 t_3e^{-i \theta_{\bm{k},2}} & 0 \\
 0 & t_3\\
\end{pmatrix}.
\end{align}
Here we have used a shorthand notation, $\theta_{\bm{k},i}:= \bm{k}\cdot \bm{a}_i$ ($i=1,2$).
We also note that $\mathcal{O}_{n,m}$ stands for the $n \times m$ zero matrix.
The band structure of this model is obtained by solving the eigenvalue problem 
$\mathcal{H}_{\bm{k}} \bm{\psi}_{\bm{k}} =E_{\bm{k}} \bm{\psi}_{\bm{k}} $,
where $E_{\bm{k}}$ is the eigenvalue and
$ \bm{\psi}_{\bm{k}}$ is the eigenvector in the form of the 14-component column vector. 
Although we can solve the eigenvalue problem numerically, 
obtaining the analytic solutions is not easy because the matrix size of $\mathcal{H}_{\bm{k}}$ is rather large. 
Nevertheless, we can obtain the exact solutions of the flat bands for special cases. 

\section{Results \label{sec:band}}
In this section, we argue the characteristic band structures of the present model.
Specifically, we focus on the case where $t_1 = t_2 = -1$ and $t_3/t_1 \in [0,1]$ is the varying parameter.
In Fig.~\ref{fig:band}, we plot the band structures for several values of $t_3/t_1$. 
We see the exact flat bands in two panels, namely, Fig.~\ref{fig:band}(a) (i.e., $t_3/t_1 = 0$) and 
Fig.~\ref{fig:band}(f) (i.e., $t_3/t_1 = 1$).  
In the following, we will explain two flat-band limits in detail. 

\subsection{Flat-band solutions for two flat band limits}
For $t_3/t_1 = 0$ [Fig.~\ref{fig:band}(a)],
the zero energy flat band is three-fold degenerate.
In fact, two of these states are trivial, since sublattices 13 and 14 become isolated sites for $t_3 = 0$.
The remaining zero-energy flat band can be obtained analytically.
A key finding is that $T_1$ and $T_2$ of Eqs.~(\ref{eq:T1}) and (\ref{eq:T2}) have common kernel, $\bm{u} =(1,-1,1,-1,1,-1)^{\rm T}$.
Namely, $T_1\bm{u} = T_2 \bm{u} = 0 $ holds. 
Having this at hand, we find that the following vector,
\begin{align}
\bm{\psi}_{\bm{k}} 
= \frac{1}{\sqrt{6}} 
\left(0,0,0,0,0,0, 1,-1,1,-1,1,-1,0,0
\right)^{\rm T}, \label{eq:fbwf_cls}
\end{align}
satisfies $\mathcal{H}_{\bm{k}} \bm{\psi}_{\bm{k}}  = 0$ for $t_3 = 0$.
This wave function has finite amplitudes on the sites at the edges of the hexagon (sublattices 7-12),
and has vanishing amplitudes the sites on the vertices of the hexagon (1-6),
which clearly indicates that the flat band arises from the destructive interference effect. 
It is also worth noting that this zero-energy flat band is robust against changing $t_2/t_1$, as far as $t_3 = 0$ is satisfied. 
Note also that the zero-energy flat band can be understood in terms of the molecular-orbital representation~\cite{Hatsugai2011,Mizoguchi2019,Mizoguchi2021_skagome}.
See Appendix~\ref{app:mo} for details.

Next, let us consider $t_3/t_1 = 1$ [Fig.~\ref{fig:band}(f)]. 
In this case, we see two exact flat bands, 
both of which have non-zero eigenenergy.
Clearly, these flat bands do not belong to conventional 
classes arising from the bipartite or line-graph-type lattice structures.
Rather, the flat bands in this case realize accidentally, meaning 
that the clear guiding principle for deriving flat bands is lacking. 
Nevertheless, we find the exact flat-band solutions 
in this case, whose derivation is elaborated in Appendix~\ref{app:FB_exact}.
We find that, 
the flat-band energies are given as,
\begin{align}
E_{\pm} = 1\pm \sqrt{2}, \label{eq:FB_en}
\end{align}
and corresponding wave functions are
\begin{widetext}
\begin{align}
\bm{\psi}_{\bm{k},\pm} 
= \frac{1}{\mathcal{N}_{\bm{k},\pm }} 
\left( 
\xi_{\bm{k},\pm},
\eta_{\bm{k},\pm},
\zeta_{\bm{k},\pm},
\eta^\ast_{\bm{k},\pm},
\zeta^\ast_{\bm{k},\pm},
\xi^\ast_{\bm{k},\pm},
1,
1,
1,
1,
1,
1,
-\frac{X_{\bm{k}}}{E_{\pm}},
-\frac{X^\ast_{\bm{k}}}{E_{\pm}}
\right)^{\rm T}, \label{eq:fbwf}
\end{align}
\end{widetext}
where 
\begin{subequations}
\begin{align}
\xi_{\bm{k},\pm} = 
-\frac{2 (E_{\pm}- e^{i( \theta_{1,\bm{k}} +   \theta_{2,\bm{k}}) })}{E_{\pm}^2-1}, 
\end{align}
\begin{align}
\eta_{\bm{k},\pm} = 
-\frac{2 (E_{\pm}- e^{- i\theta_{1,\bm{k}}} )}{E_{\pm}^2-1}, 
\end{align}
\begin{align}
\zeta_{\bm{k},\pm} = 
-\frac{2 (E_{\pm}- e^{-i \theta_{2,\bm{k}}} )}{E_{\pm}^2-1}, 
\end{align}
and
\begin{align}
X_{\bm{k}}= 1+e^{-i\theta_{\bm{k},1} } + e^{i\theta_{\bm{k},2}}. \label{eq:X}
\end{align}
\end{subequations}
$\mathcal{N}_{\bm{k},\bm{\pm}}$ is the normalization constant. 
Clearly, these wave functions are completely different from that of Eq.~(\ref{eq:fbwf_cls}).
Most remarkably, there is not sublattice that has vanishing amplitude, meaning that the
origin of the flat band cannot be attributed to the destructive interference mechanism. 

As a by-product of the exact wave functions in Eq.~(\ref{eq:fbwf}), we find that one can construct 
the topological model with keeping the exact flat bands by introducing the specific form of the imaginary next-nearest-neighbor hoppings.
The result is presented in Appendix~\ref{app:topo}.

\begin{figure*}[tb]
\begin{center}
\includegraphics[clip,width = 0.95\linewidth]{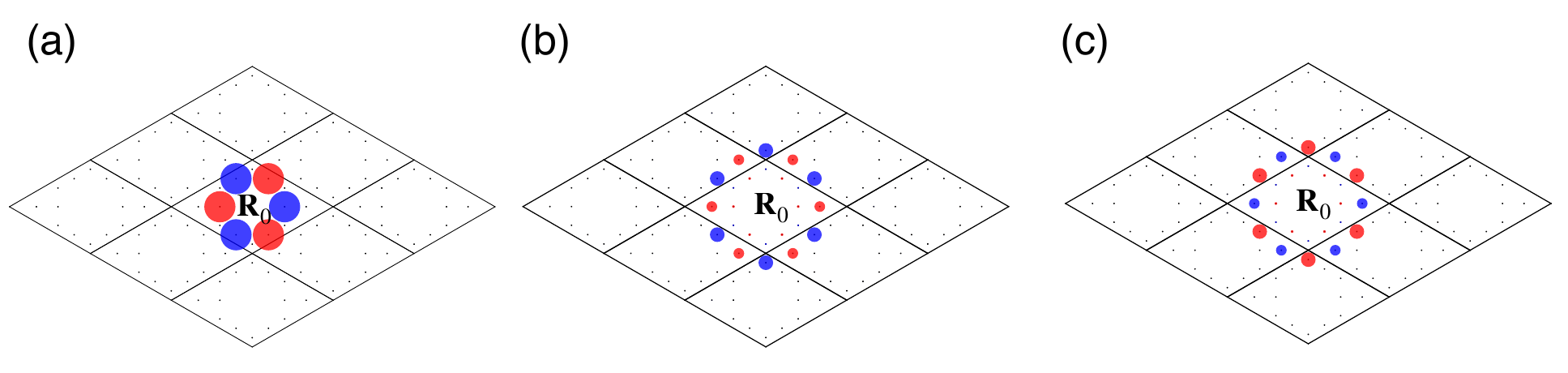}
\vspace{-10pt}
\caption{
Real-space distribution of 
(a) the CLS for $t_3/t_1 = 0$,
(b) $w_{\bm{R}_0, -} (\bm{R},\alpha)$ for $t_3/t_1 = 1$ and 
(c) $w_{\bm{R}_0, +} (\bm{R},\alpha)$ for $t_3/t_1 = 1$.
The red and blue circles indicate that the signs of the wave function are 
$+$ and $-$, respectively,
and the radii of the circles are proportional to $|w|^2$. 
}
  \label{fig:wanner}
 \end{center}
 \vspace{-10pt}
\end{figure*}
\subsection{Real-space picture of the exact flat bands}
We further address the real-space pictures of the two exact-flat-band limits. 
To this end, we derive the Wannier function.
In general, computing the Wannier function solely 
by the numerical diagonalization of the Bloch Hamiltonian
requires cautions in the gauge fixing of the wave function at each momentum.
In contrast, having the analytical solution at hand, we can straightforwardly calculate the Wannier function.
To be concrete, 
the Wannier function for the flat-band states with centered at 
the unit cell $\bm{R}_0$ is given as
 which is localized around the unit cell $\bm{R}_0$: 
 \begin{align}
w_{\bm{R}_0} (\bm{R},\alpha) 
= \frac{1}{N_{\rm u.c.}} \sum_{\bm{k}} e^{i\bm{k}\cdot(\bm{R}-\bm{R}_0)} [\bm{\psi}_{\bm{k}}]_{\alpha}, 
\label{eq:wannier}
\end{align}
with $N_{\rm u.c.}$ being the number of unit cells in the system and $\alpha$ being the sublattice index.

 We first consider the $t_3/t_1 = 0$.
 As we have mentioned, two of the flat bands are obtained by the isolated sites 13 and 14, so we consider the 
 Wannier function corresponding to the Bloch wave function of Eq.~(\ref{eq:fbwf_cls}).
 As the Bloch wave function of Eq.~(\ref{eq:fbwf_cls}) is $\bm{k}$-independent, 
 $\bm{k}$-summation in Eq.~(\ref{eq:wannier}) can be taken trivially. 
 We find that the Wannier function
 is equal to the CLS 
 that has finite amplitudes on finite number of sites and 
 zero amplitude on the remaining sites. 
This CLS is depicted in Fig.~\ref{fig:wanner}(a).
 
We next consider the case of $t_3/t_1 = 1$.
In Figs.~\ref{fig:wanner}(b) and \ref{fig:wanner}(c), we depict the real-space profile of the Wannier functions 
$w_{\pm}$ of several unit cells around the center. 
We take the summation over $\bm{k}$ in Eq.~(\ref{eq:wannier}) numerically with $72 \times 72$ meshes. 
Note that the Wannier function is real due to the symmetry, $\bm{\psi}_{-\bm{k},\pm}=\bm{\psi}_{\bm{k},\pm}^\ast$.
We see that both of the flat bands have finite amplitudes on all sublattices,
which again provides the evidence that simple interference mechanism does not apply to these flat bands.
We also see a characteristic real-space profile, namely,
the inner small hexagonal plaquette measured from the Wannier center has the small amplitudes
whereas the second hexagonal plaquette has the largest amplitudes. 

\subsection{Robust nearly flat bands}
Turing to Fig.~\ref{fig:band}, we find another important feature.
Namely, even away from the exact flat band limits [Figs.~\ref{fig:band}(b)-(e)], 
there is a reasonably flat band near the Fermi energy of the half-filled case.
This indicates that strongly-correlated physics survive even away from the exact flat-band limits.
Hence, the fused pentagon network is a suitable platform for questing flat-band-induced exotic physics. 

\section{Summary and discussions \label{sec:summary}}
We have studied the band structure of the tight-binding model of the two-dimensional fused pentagon network.
The model is remarkable in that the there are two flat-band limits upon changing the hopping parameters.
For one limit ($t_3/t_1 =0 $), the flat band is at the zero-energy and its real-space picture is given by the CLS. 
For another limit ($t_3/t_1 = 1 $), the flat bands, appearing in two different energies, realize rather accidentally.
The resulting Wannier functions are completely different from the CLS of $t_3/t_1 =0$.
Additionally, upon interpolating these two limits, the bands near the Fermi energy of the half-filled case
 remain reasonably flat (though it is not exactly flat),
which indicates that the strongly-correlated physics due to the nearly-flat bands,
such as the emergence of spin-polarized state, is expected throughout the parameter region investigated in this paper. 

As for the materials realization,
as we have mentioned, the lattice structure is inspired by the prior work on the carbon network~\cite{Maruyama2013},
Actually, the band structure obtained by the first-principles calculation is similar to that of the tight-binding model.
In the real materials, the fine-tuned conditions for the hopping parameters are not satisfied in general,
as the bond length of the black, red and green bonds in Fig.~\ref{fig:lattice} are different among each other.
Nevertheless, the fact that the nearly flat band is robust against changing $t_3/t_1$ indicates that the present carbon network 
is a suitable candidate for pursuing the flat-band-induced exotic phenomena. 
Additionally, based on this material, one can make the situation similar to $t_3/t_1 = 0$ 
by the hydrogen adsorption on the sublattices 13 and 14. 
Specifically, $\pi$-electrons do not live on the adsorbed sites hence the remaining 
network corresponds to the case of $t_3/t_1 = 0$.
Our theoretical results predict that this adsorption changes the nature of the flat band drastically. 

Finally, we address the implication of the exact flat bands for $t_3/t_1 = 1$ in the completely different physics,
namely, the interacting spin model on the same lattice with the antiferromagnetic coupling.
Indeed, in the context of the spin model, the flat band indicates the macroscopically-degenerate magnetic ground state
at the classical level~\cite{Luttinger1946,Reimers1991,Garanin1999,Isakov2004,Essafi2017,Mizoguchi2018_spin}.
In the present model, we see the exact flat band at on the top for $t_1 = t_2 = t_3 = -1$,
meaning that the flat band is on the bottom when reversing the sign of the hoppings, i.e., for $t_1 = t_2 = t_3 = 1$. 
This indicates  
the macroscopically-degenerate magnetic ground state for 
the spin model with the antiferromagnetic coupling, so
revealing its ground state will be an interesting future problem. 

\acknowledgements
This work is supported by JSPS KAKENHI
Grant No.~JP20H05664, 
JP21K14484,
JP21H05232,
JP21H05233,
JP22H00283,
JP23H05469,
JP23K03243,
and JP23K25788.
It is also supported by 
JST-CREST Grant No. JPMJCR19T1.

\appendix
\section{Molecular-orbital representation for $t_3=0$ \label{app:mo}}
In this appendix, we present the insight into the zero-energy flat band for $t_3=0$
from the molecular-orbital representation~\cite{Hatsugai2011,Mizoguchi2019,Mizoguchi2021_skagome},
by which we clarify the origin of the flat band from the linear-algebraic point of view.
The generic framework of the molecular representation is that the zero-energy flat band appears
when the $n \times n$ Bloch Hamiltonian $\mathcal{H}_{\bm{k}}$ can be represented as
\begin{align}
\mathcal{H}_{\bm{k}} = \Phi_{\bm{k}} h_{\bm{k} }\Phi^\dagger_{\bm{k}}, \label{eq:mo_rep}
\end{align}
where $\Phi_{\bm{k}}$ is an $n \times m$ ($m<n$) matrix and $ h_{\bm{k} }$ is an $m \times m$ Hermitian matrix. 
Note that $\Phi_{\bm{k}}$ can be given by the $m$ sets of the column vector, 
each of which can be regarded as the (unnormalized) wave function of the molecular orbital:
\begin{align}
\Phi_{\bm{k}} = 
\begin{pmatrix}
\bm{\phi}_{\bm{k},1} & \cdots & \bm{\phi}_{\bm{k},m}
\end{pmatrix}.
\end{align}
The flat band has (at least) $(n-m)$-fold degeneracy since there exist at least $(n-m)$ vectors $\bm{\varphi}_{\bm{k},\ell}$
satisfying $\Phi^\dagger_{\bm{k}} \bm{\varphi}_{\bm{k},\ell} = 0$ due to the fact that $\Phi^\dagger_{\bm{k}}$ is the non-square matrix
whose number of row is smaller than that of the column. 

Let us now consider the case of $t_3 = 0$ of the present model.
We omit the sublattices 13 and 14 since they are the isolated sites providing trivial zero-energy flat band.
Then the remaining $12 \times 12 $ block of the Bloch Hamiltonian is given as 
\begin{align}
\mathcal{H}^\prime_{\bm{k}} = \begin{pmatrix}
\mathcal{O}_{3,3} & S_{\bm{k}} & T_1 \\
S_{\bm{k}}^\dagger &\mathcal{O}_{3,3} & T_2 \\
T_1^\dagger & T_2^\dagger & \mathcal{O}_{6,6}  \\
\end{pmatrix}. \label{eq:ham_prime}
\end{align}

We now show that $\mathcal{H}^\prime_{\bm{k}}$ can be expressed in the form of Eq.~(\ref{eq:mo_rep}).
To be more specific, we show the explicit form of 11 molecular orbitals by which we can rewrite the Hamiltonian. 
To this end, we utilize the idea which we have used in Ref.~\onlinecite{Mizoguchi2021_skagome}.
Namely, as a first step, we define 12 molecular orbitals rather than 11, one of which is turned out to be redundant. 
Specifically, we define 
\begin{align}
\bm{\phi}_{\bm{k},j} = \bm{e}_j, 
\end{align}
 for $j=1, \cdots, 6$
 and
 \begin{subequations}
\begin{align}
\bm{\phi}_{\bm{k},7} = \bm{e}_7 + \bm{e}_8,
\end{align}
\begin{align}
\bm{\phi}_{\bm{k},8} = \bm{e}_8 + \bm{e}_9,
\end{align}
\begin{align}
\bm{\phi}_{\bm{k},9} = \bm{e}_9 + \bm{e}_{10},
\end{align}
\begin{align}
\bm{\phi}_{\bm{k},10} = \bm{e}_{10} + \bm{e}_{11},
\end{align}
\begin{align}
\bm{\phi}_{\bm{k},11} = \bm{e}_{11} + \bm{e}_{12},
\end{align}
and
\begin{align}
\bm{\phi}_{\bm{k},12} = \bm{e}_{12} + \bm{e}_7.
\end{align}
\end{subequations}
Here, $\bm{e}_j$ is the unit vector whose $j$th component is one and all the other components are zero.
Defining $\tilde{\Phi}_{\bm{k}} = \begin{pmatrix}
\bm{\phi}_{\bm{k},1} & \cdots & \bm{\phi}_{\bm{k},12}
\end{pmatrix}$,
we can rewrite $\mathcal{H}_{\bm{k}}^\prime$ as
\begin{align}
\mathcal{H}_{\bm{k}}^\prime = \tilde{\Phi}_{\bm{k}}  \tilde{h}_{\bm{k} } \tilde{\Phi}^\dagger_{\bm{k}} ,
\end{align} 
where $ \tilde{h}_{\bm{k}}$ is the $12 \times 12$ matrix
\begin{align} 
 \tilde{h}_{\bm{k}} = 
 \begin{pmatrix}
\mathcal{O}_{3,3} & S_{\bm{k}} & T^\prime_1 \\
S_{\bm{k}}^\dagger &\mathcal{O}_{3,3} & T^\prime_2 \\
T_1^{\prime \dagger} & T_2^{\prime \dagger} & \mathcal{O}_{6,6}  \\
\end{pmatrix},
 \end{align} 
with
$ S_{\bm{k}} $ is the same as that of Eq.~(\ref{eq:smat}) and 
\begin{align}
 T^\prime_1 = 
 \begin{pmatrix}
 t_1 & 0 & 0 & 0 & 0 & 0 \\
 0 & 0 & t_1 & 0 & 0 & 0 \\
  0 & 0 &0 & 0 &  t_1& 0 \\
 \end{pmatrix},
\end{align}
\begin{align}
 T^\prime_2 = 
 \begin{pmatrix}
 0 & 0 & 0 & 0 & 0 & t_1 \\
 0 & t_1 & 0 & 0 & 0 & 0 \\
  0 & 0 &0 & t_1 &  0& 0 \\
 \end{pmatrix}.
\end{align}
The key step to obtain the final form of the molecular-orbital representation is to 
notice that $\bm{\phi}_{\bm{k},7}$-$\bm{\phi}_{\bm{k},12}$ are not linearly independent with each other~\cite{Mizoguchi2021_skagome}.
Rather, they satisfy 
\begin{align}
\bm{\phi}_{\bm{k},12} = (\bm{\phi}_{\bm{k},7} + \bm{\phi}_{\bm{k},9} + \bm{\phi}_{\bm{k},11}) - (\bm{\phi}_{\bm{k},8} + \bm{\phi}_{\bm{k},10}).
\end{align}
Using this relation we can rewrite $\mathcal{H}^\prime_{\bm{k}}$ in the form of Eq.~(\ref{eq:mo_rep}).
Specifically, 
adopting $\Phi_{\bm{k}} = 
\begin{pmatrix}
\bm{\phi}_{\bm{k},1} & \cdots & \bm{\phi}_{\bm{k},11}
\end{pmatrix}$,
we set $h_{\bm{k}}$ as
\begin{align}
h_{\bm{k}} = 
 \begin{pmatrix}
\mathcal{O}_{3,3} & S_{\bm{k}} & T^{\prime \prime}_1 \\
S_{\bm{k}}^\dagger &\mathcal{O}_{3,3} & T^{\prime \prime}_2 \\
T_1^{\prime  \prime\dagger} & T_2^{\prime \prime \dagger} & \mathcal{O}_{5,5}  \\
\end{pmatrix},
\end{align}
where $T^{\prime \prime}_1 $ and $T^{\prime \prime}_2 $ are the $3 \times 5$ matrices:
\begin{align}
 T^{\prime \prime}_1 = 
 \begin{pmatrix}
 t_1 & 0 & 0 & 0 & 0  \\
 0 & 0 & t_1 & 0 & 0  \\
  0 & 0 &0 & 0 &  t_1 \\
 \end{pmatrix},
\end{align}
\begin{align}
 T^{\prime \prime} _2 = 
 \begin{pmatrix}
 t_1& -t_1 & t_1 & -t_1 & t_1  \\
 0 & t_1 & 0 & 0 & 0 \\
  0 & 0 &0 & t_1 &  0 \\
 \end{pmatrix}.
\end{align}
It is again worth mentioning that this molecular-orbital representation 
holds for any $t_1$ and $t_2$,
meaning that the zero-energy flat band exists for any $t_1$ and $t_2$.

\section{Analytic solution of the flat band for $t_1 = t_2 = t_3 $ \label{app:FB_exact}}
In this appendix, we derive  the analytic solution of the flat-band states for 
$t_1 = t_2 = t_3 = -1$. 
Unlike the case of $t_3=0$, it is not straightforward to find the molecular-orbital representation.
Hence, we explicitly solve the Schr\"{o}dinger equation 
$\mathcal{H}_{\bm{k}} \bm{\psi}_{\bm{k}} = E \bm{\psi}_{\bm{k}}$
where $\bm{\psi}_{\bm{k}} = \left( \psi_{\bm{k},1}, \cdots,  \psi_{\bm{k},14}\right)^{\rm T}$.
Hereafter, we abbreviate the ${\bm{k}}$ dependence of $\psi_{\bm{k},j}$.
A key assumption for obtaining the analytic solution is 
\begin{align}
\psi_{7} = \psi_{8} = \psi_9 = \psi_{10}=\psi_{11} = \psi_{12}.  \label{eq:assump}
\end{align}
We have found this condition by the numerical diagonalization. 
Without loss of generality, we set $\psi_{7} = \psi_{8} = \cdots, = \psi_{12} = 1$; 
the normalization of the wave function will be considered afterwards.  

Under this assumption, we solve the Schr\"{o}dinger equation explicitly.
We first focus on the sublattices $13$ and $14$, since they are connected only to the sublattices $7$-$12$.
By substituting this above assumption into the Schr\"{o}dinger equation, we have
\begin{align}
-\left(1+e^{-i\theta_1} + e^{i\theta_2}\right) =& E \psi_{13}, \notag \\
-\left(1+e^{i\theta_1} + e^{-i\theta_2}\right) =& E \psi_{14}. 
\label{eq:psi14}
\end{align}
Then we have $\psi_{13}= - \frac{X}{E}$ and $\psi_{14}= - \frac{X^\ast}{E}$ [$X$ is defined in Eq.~(\ref{eq:X})]. 
Here we assume $E \neq 0$, which is indeed the case. 

We next focus on $\psi_1, \cdots, \psi_6$, i.e., the vertexes of hexagonal rings. 
For $\psi_1$ and $\psi_6$, the Schr\"{o}dinger equation leads to
\begin{align}
\begin{pmatrix}
0 & -e^{i(\theta_1 + \theta_2)}\\
-e^{-i(\theta_1 + \theta_2)}& 0 \\ 
\end{pmatrix}
\begin{pmatrix}
\psi_1 \\
\psi_6\\ 
\end{pmatrix}
- 
\begin{pmatrix}
\psi_7 + \psi_8 \\
\psi_{10} + \psi_{11}\\ 
\end{pmatrix}
= E 
\begin{pmatrix}
\psi_1 \\
\psi_6\\ 
\end{pmatrix}.
\end{align}
From the assumption $\psi_7 = \cdots, \psi_{12} = 1$, 
we obtain 
\begin{align}
\psi_1 = \psi_6^\ast = - \frac{2[E- e^{i(\theta_1 + \theta_2)}]}{E^2-1}.  \label{eq:psi_1_6}
\end{align}
Note that we additionally assume $E\neq \pm 1$, which is also the case. 
Similarly, for $\psi_2$, $\psi_3$, $\psi_4$, and $\psi_5$, we obtain
\begin{align}
\psi_2 =&  \psi_4^\ast = - \frac{2(E- e^{-i\theta_1})}{E^2-1},  \notag \\
\psi_3  =& \psi_5^\ast = - \frac{2(E- e^{-i\theta_2} )}{E^2-1}.  \label{eq:psi_3_5}
\end{align}

Lastly, we focus on  $\psi_7, \cdots, \psi_{12}$, i.e., the sites on the edges of hexagonal rings.
We determine the eigenenergy $E$ so that the assumption of $\psi_7= \cdots = \psi_{12} = 1$ is satisfied self-consistently. 
For $\psi_7$, the Schr\"{o}dinger equation leads to
\begin{align}
-(\psi_1 + \psi_4) -e^{i\theta_1} \psi_{13} = E \psi_7. \label{eq:psi_7}
\end{align}
Substituting Eqs. (\ref{eq:psi14}), (\ref{eq:psi_1_6}), and (\ref{eq:psi_3_5}) into Eq.~(\ref{eq:psi_7}), 
as well as $\psi_7 =1$, we obtain 
\begin{align}
\frac{2}{E^2-1}(2E-\alpha) + \frac{1+\alpha}{E}= E, \label{eq:E_determine_1}
\end{align}
where we have defined $\alpha := e^{i(\theta_1 + \theta_2)} + e^{i\theta_1}$.
Equation~(\ref{eq:E_determine_1}) can be rewritten as 
\begin{align}
(E^2-2E-1) (E^2 + 2E-1-\alpha) = 0.
\end{align}
As $\alpha$ is actually the $\bm{k}$-dependent quantity,
the $\bm{k}$-dependent solutions of the above equation correspond to Eq.~(\ref{eq:FB_en}) [i.e., $E = 1 \pm \sqrt{2}$].
We note that the conditions for $\psi_8, \cdots, \psi_{12}$ lead to the same eigenenergies.
Once the eigenenergies are determined, the corresponding wave functions can obtained from 
Eqs.~(\ref{eq:assump}),
(\ref{eq:psi14}), (\ref{eq:psi_1_6}), (\ref{eq:psi_3_5}), and (\ref{eq:psi_7}).
The resulting wave functions are shown in Eq.~(\ref{eq:fbwf}).

\section{Topological flat-band model with high Chern numbers \label{app:topo}}
\begin{figure}[!htb]
\begin{center}
\includegraphics[clip,width = \linewidth]{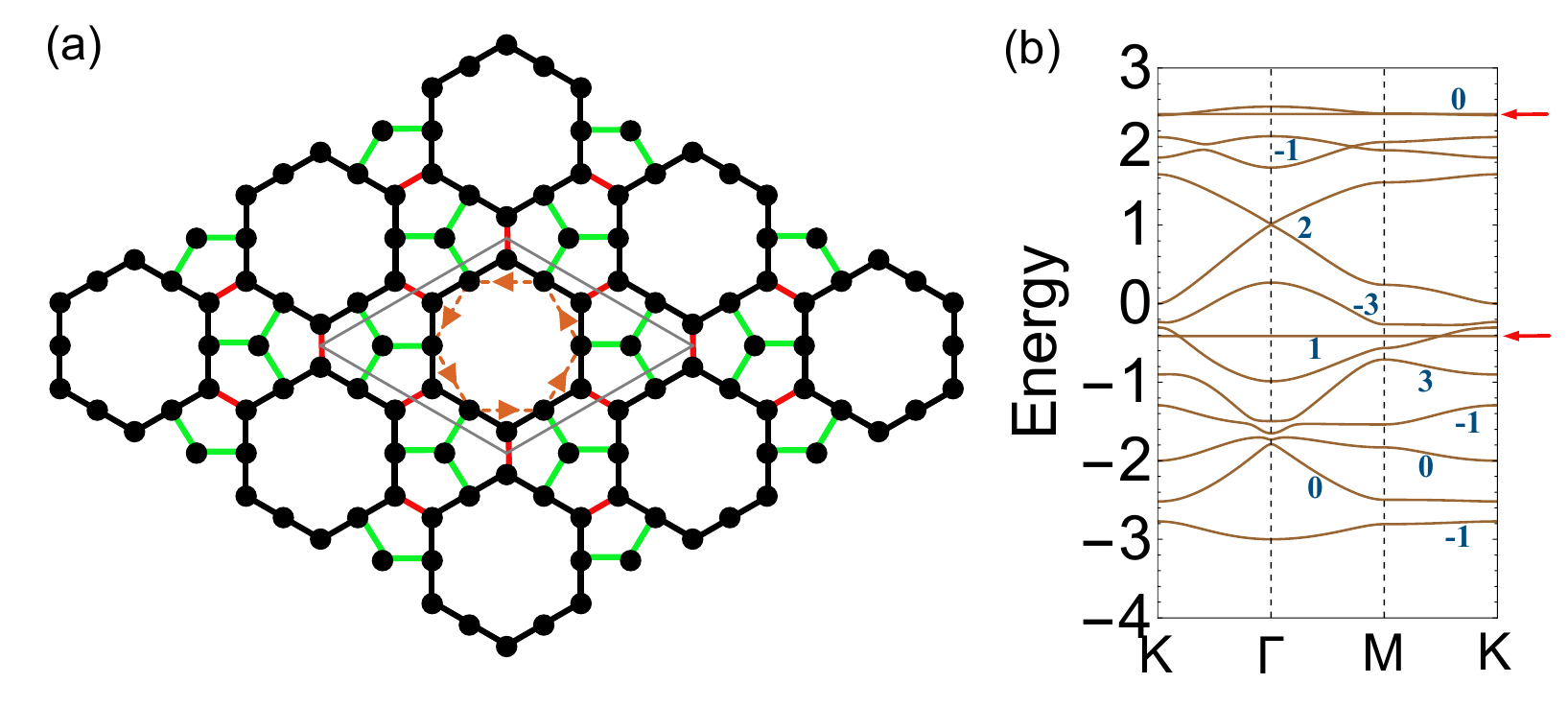}
\vspace{-10pt}
\caption{
(a) Schematic figure of additional imaginary hoppings preserving the flat bands.
The orange broken lines represent the hopping $-i\lambda$ (in the direction of the arrow). 
(b) Band structure for $(t_1,t_2,t_3,\lambda) = (-1,-1,-1,0.3)$.
Red arrows point the flat bands.
}
  \label{fig:SOI}
 \end{center}
 \vspace{-10pt}
\end{figure}
Here we present the construction of the topological flat-band model based on the present lattice structure, 
which is inspired by the analytic wave function
In fact, the flat-band wave function of Eq.~(\ref{eq:fbwf}) tells us 
what kind of additional hopping terms can preserve the exact flat bands. 
To be concrete, we introduce the pure-imaginary hoppings 
among the sites 7-12, as shown in Fig.~\ref{fig:SOI}(a).
This type of imaginary hoppings can be a source of topologically nontrivial states,
as proposed in the seminal work by Haldane~\cite{Haldane1988}. 
The band structure is shown in Fig.~\ref{fig:SOI}(b).
We also compute the Chern number of each band (or sets of bands if they are degenerate) numerically~\cite{Fukui2005}.
We see the exact flat bands are unchanged.
This is due to the fact that $\psi_7 = \cdots, \psi_{12}$ 
holds for every $\bm{k}$ and thus the additional pure-imaginary hoppings 
vanish when acting on the flat-band wave function. 
In the present set of parameters, 
the set of bands containing the flat band 
$E_-$ has the finite Chern number, whereas that containing $E_+$ has the vanishing 
Chern number.
We also see that some of the bands has the high Chern number (i.e., the absolute value of the Chern number is larger than 1),
which has been of interest recently~\cite{Wang2011,Sticlet2012,Yang2012,Wang2012_2,Trescher2012}.

\bibliographystyle{apsrev4-2}
\bibliography{fused_pentagon}
\end{document}